\documentclass[aps,pra, twocolumn, nofootinbib, showpacs]{revtex4-1} 
\usepackage[unicode=true,pdfusetitle,
bookmarks=true,bookmarksnumbered=false,bookmarksopen=false,
breaklinks=false,pdfborder={0 0 0},backref=false,colorlinks=false] {hyperref}
\hypersetup{ colorlinks,linkcolor=myurlcolor,citecolor=red,urlcolor=myurlcolor}
\usepackage{braket,colortbl,amsthm,amsmath,cleveref,amssymb,txfonts}\definecolor{myurlcolor}{rgb}{0,0,0.7}
\usepackage{graphics,graphicx}
\usepackage{color}
\usepackage{amssymb}
\usepackage{amsthm}
\usepackage{amsfonts}
\usepackage{float}
\usepackage{graphicx}
\usepackage[format = plain,labelfont = bf,up, textfont = normal , up, justification
=raggedright, singlelinecheck =false]{caption}
\usepackage{subcaption}
\usepackage{tabularx}
\usepackage{amsmath}
\usepackage{braket}

\usepackage{graphicx}
\usepackage[utf8x]{inputenc}
\usepackage{color,soul}
\usepackage{amsmath}
\usepackage{braket}
\usepackage{latexsym}
\usepackage{amssymb}
\usepackage{amsthm}
\usepackage{xcolor}
\usepackage{bm}
\usepackage{graphics,epstopdf}
\usepackage{color}\usepackage{amsmath}
\usepackage{enumitem}
\usepackage{fmtcount}
\usepackage{booktabs}
\usepackage{hyperref}
\usepackage{epsfig}

\theoremstyle{plain}

\begin{document}
	

		\title{Quantum coherence with incomplete set of pointers and corresponding wave-particle duality}

	\author{Ingita Banerjee, Kornikar Sen, Chirag Srivastava, Ujjwal Sen}
	
	\affiliation{Harish-Chandra Research Institute, HBNI, Chhatnag Road, Jhunsi, Allahabad 211 019, India}
\begin{abstract}
     Quantum coherence quantifies the amount of superposition in a quantum system, and is the reason and resource behind several phenomena and technologies. It depends on the natural basis in which the quantum state of the system is expressed, which in turn hinges on the physical set-up being analyzed and utilized. 
     While quantum coherence has hitherto been conceptualized by employing different categories of complete bases, there do exist interesting physical situations, where the natural basis is an incomplete one, an example being an interferometric set-up with the observer controlling only a certain fraction of all the slits. We introduce a quantification of quantum coherence  with respect to an arbitrary incomplete basis for general quantum states, and develop the corresponding resource theory, identifying the free states and operations. Moreover, we obtain a complementarity relation between the so-defined quantum coherence and the which-path information in an interferometric set-up with several slits, of which only a section is in control of the observer or is accessible to her. This therefore provides us with another face of the wave-particle duality in quantum systems, demonstrating that the complementarity is functional in more general set-ups than thus far considered. 
 %
  %
\end{abstract}
	\maketitle
\section{Introduction} The superposition principle, arguably the most striking property of the quantum formalism, helps in describing the resources necessary for a quantum technician to possibly beat classical devices. 
A quantum system which is in a superposition of certain (pure) quantum states is one that exists ``coherently'' in all the states. 
This superposition of quantum states, also known as \emph{quantum coherence},  can be employed in quantum devices to gain advantage over their classical counterparts, along with playing an important role in understanding important physical phenomena.



Contrary to the \emph{concept} of quantum coherence, attempts to \emph{quantify} it are relatively recent
\cite{coh1,coh2}. Its formal quantification was followed by a significant body of work. See e.g. 
\cite{coh3,coh4,coh5,coh6,coh7,coh8,coh9,coh10,coh11, 
POVM1,POVM2,POVM3,POVM4, nonorthogonal1,nonorthogonal2, lin_dep}.
The concept of quantum coherence has been found to have applications in 
quantum thermodynamics \cite{thermo1,thermo2,thermo3,thermo4,thermo5,thermo6,thermo7,thermo8,thermo9,thermo10,thermo11,thermo12,thermo13}, quantum channel discrimination \cite{qcd1,qcd2}, quantum algorithms \cite{qc1,qc2,qc3,qc4}, etc. There also exist various studies establishing connections between quantum coherence and other resources, e.g., with non-Markovianity \cite{Non-Mark1,Non-Mark2,Non-Mark3,Non-Mark4,Non-Mark5,Non-Mark6,Non-Mark7,Non-Mark8}, nonlocality \cite{nonlocal1,nonlocal2,nonlocal3}, entanglement \cite{ent1, ent2, ent3, entang1,entang2,entang3,entang4,entang5,entang6,entang7,entang8,entang9, pakhi}, and quantum discord \cite{disco1, disco2, discord1,discord2,discord3}.

Resource theories of quantum coherence are generally considered with respect to a set of states of the relevant Hilbert space that is deemed as ``natural'' to the situation under study. 
For example, in a Young's double-slit experiment, it is natural to consider the quantum coherence of the state just before impinging on the screen in the complete orthonormal basis formed by the states corresponding to the two slits. 
Early works on conceptualizations of quantum coherence focused attention on defining the same with respect to complete orthonormal bases, where ``completeness'' implies that the corresponding set of states spans the Hilbert space under consideration. Later, it was realized that it is physically reasonable to consider quantum coherence with respect to complete linearly independent bases, which could in general contain nonorthogonal states~\cite{nonorthogonal1,nonorthogonal2},
and with respect to complete linearly dependent bases~\cite{lin_dep}. It may be noted that while a set of mutually orthogonal states can be distinguished in principle, by measuring onto the states of the set, a set of linearly independent states can only be probabilistically distinguished~\cite{linearindependence1,linearindependence2}. A set of linearly dependent states cannot even be probabilistically distinguished~\cite{linearindependence1,linearindependence2}.
It is also interesting to remark that while mutual orthogonality of a set of states is a property of pairs of elements of the set - a ``local'' property, linear independence and dependence are properties of the entire set - a ``global'' property.
It may also be mentioned that a linearly dependent set is ``over-complete'' in its span, i.e., it contains more state vectors than is required to span the space. 
%
Such nonorthogonal bases appear naturally in considerations of ``coherent'' states in quantum optics~\cite{Wolf}, and in the resource theory of ``magic''~\cite{magic1,magic2.1, magic2.2, magic2.3, magic2.4,magic2.5},
with the latter being relevant in the context of resources necessary in quantum computation circuits.

It is useful to mention here that the word, ``basis'', is used in the literature of quantum information in particular and quantum mechanics in general, in a different (or rather more general) sense than in mathematical monographs. In the latter, a basis of a linear space is defined to be a set of states that are linearly independent and complete. The over-complete ``basis'' of ``coherent'' states of an electromagnetic field mode~\cite{Wolf} is an example of departure from this definition. The concepts of unextendible and uncompletable product and entangled bases (see e.g.~\cite{boundent1,boundent2,entbases1,entbases2}) provide another example of the departure, where the ``bases'' are incomplete. Yet another example is provided by the concept of quantum coherence with linearly dependent ``bases'' and the identification of magic as a quantum coherence with respect to  dependent ``bases'' ~\cite{lin_dep}. 

Quantum coherence has hitherto been analyzed in the literature with respect to complete bases. 
The concepts of unextendible and uncompletable product and entangled bases indicate that there are useful \emph{incomplete} bases in quantum information. Incomplete bases may also appear in generic experimental set-ups. 
For example, this situation may appear in a multi-slit interference experiment in which a part of the entire set of slits are 
known  to the observer, while the others are 
not, 
due to the presence of noise or leakage at those slits, or because those are for some reason not under the control of the observer or inaccessible to her. 

%
%
%

In such a 
situation, we would be forced to consider a quantum state at the output (e.g., the state just before the screen of a multi-slit interference set-up, part of whose slits are inaccessible), which is to be decomposed 
in a complete basis with some known states and some unknown ones.
The question we ask is whether we can define a concept of quantum coherence in  physical set-ups where the natural basis (the ``pointer'' states) is incomplete, i.e., it does not span the relevant Hilbert space.

 We note here that leakages or noise at slits, in general, may require the slit states to be considered as nonpure, but for simplicity, we consider them to be pure here. Further, we assume that the set of known and unknown states form a complete mutually orthonormal basis. The orthogonality of the entire set is assumed  for simplicity, but also to clearly understand the response of incompleteness in the basis on the theory and applications of quantum coherence without additional effects due to nonorthogonality (with or without linear dependence).

In this article, we have two main objectives. 
Firstly, we wish to develop a resource theory of quantum coherence with respect to an arbitrary incomplete basis. 
We will
begin by identifying the set of incoherent states, i.e., the free states of the resource theory. We provide a complete characterization of the same.
We follow this up with a definition of incoherent operations, viz., operations that does not create resourceful states out of free ones, and these will be the free operations of the resource theory. 
We present two important classes of free operations. We will then identify the measures that quantify the resource, and we consider two families of such measures, of which one does not have a parallel in the resource theories known in the literature on entanglement and quantum coherence with complete bases. We also discuss about the monotonicity properties of the measures under incoherent quantum maps. 

Secondly, we wish to identify a wave-particle duality in an interferometric set-up in which quantum coherence with respect to an incomplete basis is naturally a measure of waveness. We will measure the particle nature by using the which-path information of the detectors in the set-up.

The remainder of the paper is organized in the following manner. We frame the resource theory by first introducing the structure of the incoherent states and incoherent operations  with respect to an arbitrary incomplete basis in Sec. \ref{sec1}. Then in Sec. \ref{sec2}, we establish the quantifiers of the coherence.
We show the monotonicity of a distance-based coherence measure under incoherent quantum operations in the same section. 
In the next section, viz. Sec. \ref{sec3}, we consider a four-slit interferometric set-up, in which the observer has control of only two slits. A wave-particle duality relation is obtained in this set-up where the waveness is quantified by the quantum coherence with respect to an incomplete basis, viz. the set consisting of the states corresponding to the two slits on which the observer has control over. 
We present a conclusion in Sec.~\ref{sec4}.
 
 \section{Incoherent states and incoherent operations for an incomplete basis} \label{sec1}

In any resource theory, it is important to identify which states of the corresponding are ``free'' and which are ``resourceful'', and a similar categorization for the available operations. The canonical example of a resource theory is that of entanglement of shared systems, in which case, the usual set of free states are the unentangled or separable states, while the entangled states are the resourceful ones. 
In the same resource theory, the set of free operations is often identified as the class of local quantum operations and classical communication, although the class of separable superoperators (see \cite{nonlocal} in this regard) and those that preserve the positivity of partial transpose~\cite{Per-Hor1,Per-Hor2} are also contenders. In a resource theory of quantum coherence, free states are the  ``incoherent'' states, whose meaning shifts with the flavor of quantum coherence we wish to pin down. 
The ``standard'' quantum coherence is defined with respect to a fixed complete set of orthonormal (pure) states~\cite{coh1, coh2},  forming a complete orthonormal basis of the relevant Hilbert space, and the corresponding set of incoherent states is defined as the one consisting of states that are  diagonal, when expressed in that basis. The motivation comes from the fact that the density matrices for pure states have off-diagonal elements with respect to a complete orthogonal basis, when the matrix is written in that basis, if and only if there exist a non-trivial superposition of two or more states of that basis in the pure state. 
This idea is then carried over to other resource theories of quantum coherence, where more general ``bases'' are considered which may contain nonorthogonal states~\cite{nonorthogonal1, nonorthogonal2} and even be over-complete~\cite{lin_dep}. 
 
 In this section, we begin by formally defining  an incomplete basis. We then motivate the definition of incoherent states with respect to such an incomplete basis. Subsequently, we  define and present structures of incoherent operations in a resource theory of quantum coherence with respect to the same basis. 
\\  

\subsection{Incomplete basis} Let $B_I=\{ |0 \rangle, |1\rangle, \ldots, |n-1\rangle\}$ be a \emph{fixed} set of pure states in a Hilbert space \(H\), that does not span \(H\).   And let $B^c_I=\{ |n \rangle, |n+1\rangle, \ldots, |d-1\rangle\}$ be  \emph{any} disjoint set of states (i.e.,   \(B_I \cap B^c_I = \emptyset\)), also from \(H\), 
such that $B_I \cup B^c_I$ forms a complete basis of \(H\), i.e., \(\mbox{Span} (B_I \cup B^c_I) = H\). The set $B_I$ is fixed, whereas $B^c_I$ is variable, and  the resource theory of quantum coherence that we wish to consider is with respect to $B_I$, an incomplete basis, because \(\mbox{Span} (B_I) \subsetneq H\). 
In this article, we only consider mutually orthogonal states in sets $B_I \cup B^c_I$, to clearly understand the response of the resource theory to a change in the basis from complete to incomplete, without additional effects due to nonorthogonality or linear dependence. Therefore, \(n < d = \dim H\). Note that if the cardinality, \(n\), of \(B_I\) is only one less than the dimension, \(d\), of \(H\), then \(B_I^c\) is unique up to a phase in the constituent state. Note also that while \(B_I^c\) is a variable for a given \(B_I\), its span, \(\mbox{Span}(B_I^c)\), is a fixed space. \\

  
\subsection{Incoherent states} 
In the following, we will refer to the resource theory of quantum coherence with respect to a complete orthonormal set as the ``standard'' resource theory of quantum coherence. To identify the incoherent states in our resource theory of quantum coherence with respect to an incomplete orthonormal set, \(B_I\), we consider all states that are incoherent in the standard theory for at least one extension of \(B_I\) to \(B_I \cup B_I^c\). 
This parallels the definition of a separable state of a shared system, which is defined as one which can be written as a probabilistic mixture of pure separable states of the same system for at least one such decomposition~\cite{eprcorr}. Such a scheme underlines the resource perspective of the corresponding definition. 
%
%
This principle leads us to the following structure of incoherent states with respect to the fixed incomplete basis, \(B_I = \{|i\rangle\}_{i=0}^{n-1}\): 
	\begin{equation}
			\rho_I=q\left(\sum_{i=0}^{n-1} p_i |i\rangle\langle i|\right)\oplus (1-q)\rho_{d-n}, \label{eq5}
		\end{equation}
		where 
	\begin{itemize}
		   \item \(p_i \geq 0\), $\sum_i p_i=1$, 
		   and  $0\leq q\leq 1$, and
	%
	\item $\rho_{d-n}$ is an arbitrary  density matrix on \(\mbox{Span}(B_I^c)  \equiv \mathbb{C}^{d-n}\).
	\end{itemize}
		These are therefore the ``free states'' of the resource theory of quantum coherence  in an incomplete basis. Let us denote the set of all free states corresponding to an incomplete basis \(B_I\) as \(\mathcal{F}_{B_I}\). Note that while the density matrix, \(\rho_{d-n}\), is supported on \(\mbox{Span}(B_I^c)\), the density,
	\(\sum_{i=0}^{n-1} p_i |i\rangle \langle i|\), is supported on \(\mbox{Span}(B_I) \equiv \mathbb{C}^n\).
	Note also that	
	since the $\rho_{d-n}$ is supported on $\mbox{Span}(B^c_I)$, and the set $B^c_I$ can be varied arbitrarily within \( \mbox{Span}(B_I^c)\) while preserving orthonormality, there will always exist one such set so that $\rho_{d-n}$ is diagonal when expressed as a matrix in that basis of \(\mbox{Span}(B_I^c)\). It is easy to see that the set \(\mathcal{F}_{B_I}\) is  convex in the space of all density matrices. It is also closed in the space of density operators with respect to any reasonable distance measure in that space.\\
	

\subsection{Incoherent operations} Along with free and resourceful states, it is also important in a resource theory to identify the free and resourceful operations. In the context of resource theories of quantum coherence, the free operations are often also referred to as incoherent operations. In the resource theory of quantum coherence with respect to an  incomplete basis, we identify the incoherent operations as those that preserves the set of incoherent states for that basis. A free operation, \(\Phi_I\), in our resource theory is therefore  any 
completely positive trace preserving (CPTP) map that creates an incoherent state, \(\rho_I^{\prime} = \Phi_I(\rho_I) \in \mathcal{F}_{B_I}\), by acting on any incoherent state \(\rho_I \in \mathcal{F}_{B_I}\). Let \(\mathcal{F}_{B_I}^\mathcal{O}\) be the set of all free operations for the resource theory of quantum coherence with respect to the incomplete basis, \(B_I\).

Let \(\{K_m\}\) be the set of Kraus operators corresponding to a free map \(\Phi_I\), so that 
%
%
$\Phi_I(\rho_I)=\sum_m K_m \rho_I K_m^\dagger$, where 
$\sum_m K_m^\dagger K_m=\mathbb{I}_{d}$. We will denote  the identity operator on \(\mathbb{C}^p\) as \(\mathbb{I}_p\).
We now identify two classes of free operations. We describe them by referring to the corresponding sets of Kraus operations.  
The first class is given by maps that have the Kraus decomposition with Kraus operators of the  following form:
\begin{equation}
			K_{m}=\begin{bmatrix} 
 P_m & O_1 \\
 O_2 & Q_m \\

\end{bmatrix},
\quad
\label{eq4} 
\end{equation}
%
%
		where 
		\begin{itemize}
		    \item $P_m$ are Kraus operators such that $\sum_m P_m \rho^D P_m^\dagger$ maps any state, \(\rho^D\), that is diagonal in the basis \(B_I\) (and supported on \(\mbox{Span}(B_I)\)), to another state that is also diagonal state in the same basis (and supported on the same span). Note that $\sum_m P_m^\dagger P_m=\mathbb{I}_{n}$.  
		    \item$Q_m$  are an arbitrary set of Kraus matrices such that $\sum_m Q_m^\dagger Q_m=\mathbb{I}_{d-n}$.
		    \item
		    $O_1$ and $O_2$ represent null matrices, possibly rectangular, of appropriate dimensions.
		\end{itemize}
	The second class of free operations are given by those that have the Kraus decomposition with Kraus operators of 
	the following	form:
\begin{equation}
\begin{bmatrix} 
 O_3 & O_1 \\
 R_m & Q_m \\
\end{bmatrix},
\quad
\label{kya} 
		\end{equation}
		where 
		\begin{itemize}
		    \item $Q_m$ and $R_m$ are  arbitrary Kraus matrices of dimensions $d-n \times d-n$ and $d-n \times n$, respectively. Note that $\sum_m Q_m^\dagger Q_m=\mathbb{I}_{d-n}$ and $\sum_m R_m^\dagger R_m=\mathbb{I}_{n}$. 
		    \item
		    $O_1$ and $O_3$ represent null matrices of appropriate dimensions.
		\end{itemize}


\noindent \textbf{Proof that the two classes provide free maps:}
The free states in Eq.~\eqref{eq5} can be expressed in matrix form as
\begin{equation*}
    \rho_{I}=\begin{bmatrix}
     q\rho_{1} & O_1\\
     O_2 & (1-q)\rho_{d-n} \\
    \end{bmatrix},
    \end{equation*}
    where $\rho_{1}=\sum_{i=0}^{n-1} p_i |i\rangle\langle i|$. Therefore, for an operation of the form given in Eq.~\eqref{eq4}, we have
    \begin{eqnarray*}
       \rho_I'= \sum_m K_m\rho_{I} K_m^\dagger&=&\left[ \begin{matrix}
         P_m & O_1\\  
         O_2 & Q_m \\  
        \end{matrix}\right] \left[   \begin{matrix} q\rho_1 & O_1\\
        O_2 & (1-q)\rho_{d-n} \\
        \end{matrix} \right] \left[
        \begin{matrix}
        P_m^\dagger & O_1\\
        O_2 & Q_m^\dagger\\
          \end{matrix}\right]\nonumber\\
          &=& \sum_m
          \left[\begin{matrix}
         q P_m \rho_1 P_m^\dagger & O_1\\
          O_2 & (1-q) Q_m \rho_{d-n} Q_m^\dagger \\
          \end{matrix}\right].
          \end{eqnarray*}
			By using the properties of $P_m$ and $Q_m$, mentioned earlier, we have that  $\sum_m P_m \rho_1 P_m^\dagger$ is a diagonal density matrix supported on \(\mbox{Span}(B_I)\), while $\sum  Q_m \rho_{d-n} Q_m^\dagger$ is a density matrix supported on \(\mbox{Span}(B_I^c)\), so that
            after the operation of a map from the first class 
    on $\rho_I$, the resultant matrix, $\rho_I'$,  still has the same  structure of the free states (Eq.~(\ref{eq5})). This completes the proof for the first class of maps.
Applying an operation from the second class on an arbitrary free state, we get
\begin{eqnarray*}
               \rho_I''
               &=&\sum_m \begin{bmatrix}
           O_3 & O_2\\
           R_m & Q_m\\
           \end{bmatrix}
           \begin{bmatrix}
q\rho_1 & O_1\\
O_2 & (1-q)\rho_{d-n}\\
\end{bmatrix}
\begin{bmatrix}
 O_3 & R_m^\dagger\\
 O_1 & Q_m^\dagger\\
 \end{bmatrix}
 \nonumber \\&=& \sum_m
 \begin{bmatrix}
  O_3 & O_1\\
  O_2 & q R_m\rho_1R_m^\dagger+ (1-q) Q_m\rho_{d-n}Q_m^\dagger
   
\end{bmatrix}.
            \end{eqnarray*}
The upper block of $\rho_I''$ has no non-zero elements and the lower block is a hermitian positive matrix having unit trace. Thus $\rho_I''$ is an element of \(\mathcal{F}_{B_I}\), the set of free states.
 \hfill \(\square\)
 \vspace{0.5cm} \\    
 
  It should be noted here that Kraus operators corresponding to an arbitrary incoherent operation may not have the form given in Eqs.~\eqref{eq4}  or (\ref{kya}). But if a set of Kraus operators do have those structures, then the corresponding CPTP map will be an incoherent operator for the resource theory of quantum coherence with respect to the incomplete basis, \(B_I\).       
    
\section{Quantifying coherence for  incomplete bases} \label{sec2}

An important aspect of a resource theory is its ability to quantify the resource in a reasonable manner. 
In this section, we present 
various possible measures of quantum coherence with respect to an incomplete basis. 

A functional $C(\cdot)$ on the space of density matrices 
can be considered to be a valid measure of quantum coherence with respect to an incomplete basis, \(B_I\), of a Hilbert space, \(H\), if it satisfies the following properties for an arbitrary density matrix \(\rho\) on \(H\). 
\begin{itemize}
	\item
	\textbf{Faithfulness.} $C(\rho)$ = 0  if and only if $\rho$ is an incoherent state, i.e., iff \(\rho \in \mathcal{F}_{B_I}\). 
	
	\item
	\textbf{Monotonicity.} $C(\rho)$ must be non-increasing under any incoherent operation, i.e.,
	\begin{equation*}
C( \Phi_{I}(\rho)) \leq C(\rho) \forall \rho \mbox{ and } \forall \Phi_I \in \mathcal{F}_{B_I}^\mathcal{O}.
	\end{equation*}
\end{itemize}

\noindent\textbf{Distance-based measures.} One way of quantifying the coherence is to use distance-based measures, which are contractive under CPTP operations. Such a measure can then  be defined as 
\begin{equation*}
	C_D(\rho):=\min_{\rho_I \in \mathcal{F}_{B_I}} D(\rho,\rho_I),
\end{equation*}
where $D$ represents a distance measure on the space of density operators. The faithfulness of such measures  follows directly if  \(\mathcal{F}_{B_I}\), the set of free states, is closed in the space of density operators with respect to the distance, \(D\). 
And the monotonicity follows if the distance measure employed is contractive.


%

One such distance measure is the trace distance between two density matrices, which is defined via the  Schatten-1 norm, with the 
Schatten-$p$ norm of a matrix, \(M\), being given by
\(	||M||_p=\text{Tr}(|M|^p)^{\frac{1}{p}}\).
Thus the trace-distance quantum coherence of the state $\rho$ with respect to the incomplete basis, \(B_I\) is
	given by	\begin{equation}\label{prakriti}
	C_{tr}(\rho)=\frac{1}{N}\min_{\rho_I \in \mathcal{F}_{B_I}} \text{Tr}(|\rho-\rho_I|),
	\end{equation}
		where \(N\) is a ``normalization factor'' that depends on the dimensions of \(H\) and \(\mbox{Span}(B_I)\), and  that has been appended to the definition to keep  
		the maximum value of $C_{tr}$ as unity for all density matrices on \(H\). For example, for a quantum system corresponding to which the Hilbert space is four-dimensional, and an incomplete basis that consists of two orthonormal states, $N = \frac{4}{3}$.
		We now show that the trace-distance measure of quantum coherence with respect to an incomplete basis satisfies the monotonicity property.	
	 We know that the trace distance is contractive under CPTP maps~\cite{tr1}, i.e.,  
     \begin{equation*}
     \text{Tr}(|\rho-\rho_I|)\leq \text{Tr}(|\Phi(\rho)-\Phi(\rho_I)|)
     \end{equation*} 
     \(\forall\) CPTP maps, \(\Phi\), and for all \(\rho\) and \(\rho_I\). 
     Since incoherent operations, $\Phi_I$,  are also CPTP maps, 
     we have
     \begin{equation}
     C_{tr}(\rho) \leq  \frac{1}{N}\min_{\rho_I \in \mathcal{F}_{B_I}}\text{Tr}|\Phi_I(\rho)-\Phi_I(\rho_I)|.\nonumber \end{equation} 
     Since \(\Phi_I\) is an incoherent map, \(\Phi_I(\rho_I)\) is an incoherent state, so that 
     \begin{equation}C_{tr}(\rho)\leq \frac{1}{N}\min_{\rho_I \in \mathcal{F}_{B_I}}\text{Tr}|\Phi_I(\rho)-\rho_I|,
     \nonumber
     \end{equation} 
    where the last quantity is exactly \(C_{tr}(\Phi_I(\rho))\), so that we have the intended monotonicity relation.  
    The proof is similar for any contractive distance measure, and includes for example, the relative entropy distance~\cite{relent}. We remember, however, that the relative entropy ``distance'' is not symmetric with respect to its arguments, and does not satisfy the triangle inequality.
    \\
    
    
\noindent \textbf{Measures based on minimal completion of the incomplete basis.} 
Let us pause for a moment and look back at the resource theory 
of
entanglement, with respect to the measures used there. Distance-based measures of entanglement are of course common, an oft-used one being the relative entropy of entanglement~\cite{entan1,entan2,entan3}. Another important set of measures use the concept of convex roof for a measure known for pure states, a good example being the entanglement of formation~\cite{mix_ent}. We have already discussed how to construct distance-based measures in the resource theory under consideration. A similar set of measures in the current resource theory can be constructed by using the convex-roof approach. Both the constructions can and has been done in the literature in resource theories of quantum coherence with respect to complete bases. 

The resource theory of quantum coherence with respect to an incomplete basis, however, offers a set of measures that are unique to it, and in particular cannot possibly be conceptualized  in the resource theories of entanglement and, in principle, in those of quantum coherence with respect to complete bases. 
The quantum coherence with respect to an incomplete basis, \(B_I\), of a density matrix, \(\rho\), on a Hilbert space, \(H\), can be quantified as the minimum over all possible completions of \(B_I\) to a complete orthonormal basis of \(H\) of a quantum coherence of \(\rho\) with respect to the completed basis. Let us denote the measure by \(C_{B_I}\). Although we have ignored it in the notation, the measure also depends on the measure for the completed basis, and let us denote the measure with respect to the completed basis, \(B_I \cup B_I^c\), by \(C_{B_I \cup B_I^c}\), where the measure for the completed basis can be any valid measure of quantum coherence for a complete orthonormal basis, including, e.g., the relative entropy of quantum coherence and the  \(l_1\)-norm of quantum coherence~\cite{coh1, coh2}. Therefore, we have 
\begin{equation}
C_{B_I}(\rho)=\min_{B^c_I}C_{B_I \cup B^c_I}(\rho),
\end{equation}
    where the minimization is over all completions of the incomplete basis \(B_I\) to a complete orthonormal basis of \(H\). 
    Notice that the faithfulness property will still be satisfied by the minimal-completion measure just defined, if the same is satisfied by the ``seed'' measure, \(C_{B_I \cup B_I^c}\). 
    The monotonicity property of the minimal-completion measure is however still an open question, in general. 


    \section{Wave-particle duality} \label{sec3}
    Quantum coherence has widely been regarded as a measure of ``waveness'' of a quantum system in a certain situation. The traditional wave-particle duality relations, however, typically use a visibility-based measure of waveness (see e.g.~\cite{wp3,exp1,exp3,exp4,exp5,dual12,dual13,dual14,dual15,dual16,dual17,dual18,dual19,dual20,dual21,dual22,dual23,dual24,dual25,dual10}). It is therefore interesting to see whether the wave-particle duality relations hold even if the traditional measures of waveness are replaced by quantum coherence. This has already been attempted in the literature. See e.g.~\cite{nonorthogonal2,lin_dep,exp6,exp7,dual5,dual8,dualcoh1,dualcoh2,dualcoh3,dualcoh4,dualcoh6,dualcoh7}. 
    
    

In this section, we demonstrate a complementary relation between the wave and  particle natures of a quantum system within a four-slit interferometric setup, where we have assumed that although there are four slits, the observer has control over only a specific two of them.  This is a scenario where the wave nature of the quantum system can be estimated by its quantum coherence in an incomplete basis consisting of two orthonormal states in the four-dimensional Hilbert space describing the output of the four slits.  We quantify the corresponding particle nature  by a distinguishability measure of the paths corresponding to the slits in control of the observer.

    
     Let the four slits in the set-up be denoted by \(|0\rangle\), \(|1\rangle\), \(|2\rangle\), and \(|3\rangle\). They form a complete orthonormal basis of the physical system being considered.
     %
     %
     %
     Among them, the first two are the ones which are under the control of the observer, so that \(\{|0\rangle, |1\rangle\}\) represents the fixed incomplete basis of the system under consideration. 
    %
      The state of the system, assumed pure, can be represented as a linear combination of the complete basis states as
    \begin{equation*}
    |\psi\rangle=\alpha|0\rangle + \beta|1\rangle + \gamma|2\rangle +\delta |3\rangle.
    \end{equation*} 
   \label{6}
   Here, $|\alpha|^2$, $|\beta|^2$, $|\gamma|^2$, and $|\delta|^2$ are the probabilities that the physical entity is passing through the first, second, third, and fourth slit respectively, and $|\alpha|^2+|\beta|^2+|\gamma|^2+|\delta|^2=1$.
We assume that detectors are  placed on the path of each slit.
The initial state of the  detectors 
is $|d\rangle$. The physical system, \(D\), of the detectors then interacts with the system, \(Q\), that passes through the slits, 
and gets entangled with it. If the system, \(Q\), passes through the $i^{\textsuperscript{th}}$ slit, then after the  interaction between \(Q\) and \(D\), the detector state is represented as  $|d_{i}\rangle$ (for $i=0,1,2,3$). The interaction of the detectors with the system \(Q\) is represented by a controlled unitary transformation of the form
    \begin{equation*}
        U|i\rangle|d\rangle=|i\rangle|d_i\rangle, 
    \end{equation*}
    for $i=$ 0, 1, 2, and 3.
     The joint state of system, \(Q\),  and the detectors after this interaction is given by 
    \begin{equation*}
    |\Psi_{QD}\rangle=\alpha|0\rangle|d_0\rangle + \beta|1\rangle|d_1\rangle + \gamma|2\rangle|d_2\rangle +\delta |3\rangle|d_3\rangle.
    \end{equation*}
The $|d_{i}\rangle$ are normalized and are assumed to be linearly independent.      We will next quantify the wave  and particle natures of the physical system considered, and subsequently point to a complementary relation between them, within the set-up being analyzed. 
    \subsection{Wave nature quantification using quantum coherence in incomplete basis} \label{1}
    Our first step is to quantify the wave nature of the system being considered, and we do it by employing the trace distance measure of quantum coherence, $C_{tr}$, defined in Eq.~\eqref{prakriti}, in the incomplete basis \(\{|0\rangle, |1\rangle\}\). We evaluate the coherence in the state of the system, \(Q\),  after the interaction with the detectors. For this, we will have to retrieve the state, $\rho_Q$, from the joint state, \(|\Psi_{QD}\rangle\). Mathematically,  we obtain it by tracing out the detector's part: $\rho_Q=\text{Tr}_{D}|\Psi_{QD}\rangle\langle\Psi_{QD}|$.
    The explicit form of $\rho_Q$ is given by
    \begin{equation*}
        \rho_Q=\begin{bmatrix}
         |\alpha|^2\langle d_0|d_0 \rangle & \alpha\beta^*\langle d_1|d_0\rangle &    \alpha\gamma^*\langle d_2|d_0\rangle & \alpha\delta^*\langle d_3|d_0\rangle\\
         \alpha^*\beta\langle d_0 |d_1\rangle &
         |\beta|^2\langle d_1|d_1\rangle & \beta\gamma^*\langle d_2|d_1\rangle & 
         \beta\delta^*\langle d_3|d_1\rangle\\
         \alpha^*\gamma\langle d_0|d_2\rangle &
         \beta^*\gamma\langle d_1|d_2\rangle &
         |\gamma|^2\langle d_2|d_2\rangle &
         \gamma\delta^*\langle d_3|d_2\rangle\\
         \alpha^*\delta\langle d_0|d_3\rangle &
         \beta^*\delta\langle d_1|d_3\rangle &
         \gamma^*\delta\langle d_2|d_3\rangle &
         |\delta|^2\langle d_3|d_3\rangle
        \end{bmatrix}.
    \end{equation*}
     The form of an arbitrary incoherent state, $\rho_I$, is given in Eq. \eqref{eq5}. Here we are dealing with a Hilbert space having dimension four and the number of states in the fixed incomplete  basis states is two. Therefore, in this case, $\rho_{d-n}$ is any arbitrary density matrix on \(\mathbb{C}^2\) (see Eq. \eqref{eq5}). So, we can consider the form of a general single-qubit state on the Bloch sphere to represent $\rho_{d-n}$, which is \(\rho_2\) here:
    \begin{equation*}
        \rho_2= \frac{1}{2}(I+\vec{r}.\vec{\sigma}).
    \end{equation*} 
    Here, $\vec{r}\equiv (a,b,c)$ is any three-dimensional vector having real components, $a$, $b$, and $c$, with the constraint, $|\vec{r}| \leq 1$, and $\vec{\sigma}\equiv(\sigma_x,\sigma_y,\sigma_z)$ represents the Pauli matrices.
    Hence the form of the free state in this case is given by
    \begin{equation*}
        \rho_I= \begin{bmatrix}
         qp_1 & 0 & 0 & 0\\
         0 & qp_2 & 0 & 0\\
         0 & 0 & \frac{(1-q)(1+c)}{2} & \frac{(1-q)((a-ib)}{2}\\
         0 & 0 & \frac{(1-q)(a+ib)}{2} & \frac{(1-q)(1-c)}{2}\end{bmatrix}.
    \end{equation*}
 Next we measure the coherence of the  state  using the coherence measure given in Eq. \eqref{prakriti}, to get
   \begin{equation*}
        C_{tr}(\rho_Q)=\frac{3}{4}\left(\min_{\rho_I} \text{Tr}|\rho_Q-\rho_I|\right)=\frac{3}{4}\sum_i |\lambda_i|,
    \end{equation*}\\
    where $\lambda_i$'s are the eigenvalues of the matrix $\rho_Q-\rho_I$, for the minimal \(\rho_I\).
    The quantity $C_{tr}(\rho_Q)$ gives us the ``amount of wave nature'' that resides in the state $\rho_Q$  after the interaction of \(Q\) with the detector system.
 
    \subsection{Quantifying the particle nature: Distinguishability of paths} \label{2}
   The particle nature of the quantum system, \(Q\), is associated with our ability to detect through which slit the system, \(Q\), has passed. When we track down the path of the system, \(Q\), we are doing it at the cost of a decrease in  visibility of the interference fringes. To quantify the particle nature, we utilize the path distinguishability or the which-path information of the quantum system, \(Q\). If we had considered our detector states to be orthogonal, then each path could have been distinguished with unit probability. And in that case, quantum coherence of \(Q\) would have vanished completely. Therefore, for generality, we consider detector states which may not be orthogonal to each other, so that partial knowledge about distinguishability of the detector states, and correspondingly the 
   which-path information, can be acquired. 
   
     Distinguishing between quantum states is an essential task in quantum mechanics, 
     if not fully, then to a certain degree, with the minimum quantity of mistakes committed. Quantum state discrimination involves strategizing ways to identify quantum states from an ensemble. When we have non-orthogonal states in the ensemble, there are at least two categories of quantum state discrimination strategies, namely the minimal-error quantum state discrimination~\cite{mes1,mes2,mes3,mes4} and unambiguous quantum state discrimination (UQSD) \cite{um0,um1,um2,um3,um4,um5,um6,um7,um8,um9,um10,um11}. In the former, there is always an answer but there exists a certain probability of the answer being wrong, and the optimal strategy minimizes this probability. In the latter case, there may not always be an answer but when there is, the answer is certainly correct. In UQSD, the task is to minimize the probability of not obtaining an answer. In this work, we have employed the method of UQSD to detect the path through which the system, \(Q\), has passed.
   
    UQSD was first introduced for discrimination of two non-orthogonal states \cite{um1}. Let us suppose that we have a source that produces  two non-orthogonal states, viz. $|\psi_1\rangle$ and $|\psi_2\rangle$, with probabilities, $p_1$ and $p_2$ respectively. The probability, $P$, of unambiguously discriminating between the two states, $|\psi_1\rangle$ and $|\psi_2\rangle$, i.e., unambiguously identifying which one of the two state has been produced by the source in a particular instance, 
    is given by 
  the Ivanovic-Dieks-Peres  limit~\cite{um1, um2, um3}:
   \begin{equation}\label{kaise}
       P = 1-(\sqrt{p_1 p_2}|\langle \psi_1|\psi_2\rangle|+\sqrt{p_2 p_1}|\langle{\psi_2}|\psi_1\rangle| ).
   \end{equation}
   
  The use of UQSD has also been generalized for an arbitrary number of non-orthogonal states~\cite{um10,um11}. If we consider an ensemble of $n$ non-orthogonal but linearly independent states, $|\psi_1\rangle,|\psi_2\rangle,\ldots,|\psi_n\rangle$ 
  with probabilities $p_1,p_2,\ldots,p_n$ respectively, then an upper limit to the  probability to identify the given quantum state in a particular run of the experiment is given by \cite{um10,um11}
   \begin{equation*}
       P \leq 1-\frac{1}{n-1} \sum_{i\neq j} \sqrt{p_i p_j}|\langle{\psi_i}|\psi_j\rangle.
   \end{equation*}
   
  In the interferometric set-up that we are considering,  we need to quantify the distinguishability between the two  detector states corresponding to first and second slits, which are the only slits in control of the observer. This will be assumed to evaluate the particle nature of the system, \(Q\). We trace over the system, \(Q\), in the state, $|\Psi_{QD}\rangle$, of the joint system of \(Q\) and \(D\),
  to obtain the reduced detector state, $\rho_D$. $\rho_D$ is a mixture of the detector states, $|d_0\rangle,|d_1\rangle,|d_2\rangle,|d_3\rangle$, and is given by
   \begin{eqnarray*}
       \rho_D &=& \text{Tr}_Q |\Psi_{QD}\rangle\langle\Psi_{QD}|\\
       &=&|\alpha|^2|d_0\rangle\langle d_0| +|\beta|^2|d_1\rangle\langle d_1|+|\gamma|^2|d_2\rangle\langle d_2|+|\delta|^2|d_3\rangle\langle d_3|.
   \end{eqnarray*}
   But this is the relevant  detector state when all the slits and corresponding detectors are accessible to the observer.
   Since in our case, only the first two slits are accessible to the observer, we aim  to find an effective detector state, $\rho'_D$, of the accessible detectors.
   Consider a measurement on the detector state $\rho_D$, with the positive operator-valued measurement (POVM) elements being constituents of the set, $\{A_0,~A_1,~A_?\}$, so that $A_0 + A_1 + A_?=\mathbb{I}_4$, with 
   \begin{eqnarray}
   A_0&=&c|d^\perp_{123}\rangle\langle d^\perp_{123}|, \nonumber \\
   A_1&=&c|d^\perp_{023}\rangle\langle d^\perp_{023}|,
   \end{eqnarray}
   where $|d^\perp_{ijk}\rangle$ is chosen such that it is orthogonal to $|d_i\rangle$, $|d_j\rangle$, and $|d_k\rangle$, and \(c\) is a positive real number such that the POVM element, \(A_{?}\), is positive.
   $|d^\perp_{ijk}\rangle$ can be easily determined by employing the Gram-Schmidt orthonormalization process with the states $|d_i\rangle$, $|d_j\rangle$, and $|d_k\rangle$.
   Note that $A_0$ and $A_1$ act non-trivially only on the detector states $|d_0\rangle$ and $|d_1\rangle$ respectively, whereas $A_?$ acts non-trivially on all the detector states. In the situation when  the observer of the interferometric set-up is assumed to have control over the first and second slits only, an effective detector state is obtained by discarding the outcome of the POVM element, \(A_?\), and choosing \(\rho_D'\), within the generalized von Neumann-L{\" u}ders rule (except for a unitary freedom)~\cite{Busch},
as 
   \begin{eqnarray*}
   \rho'_D&=&\frac{A^{1/2}_0 \rho_D A^{1/2}_0 + A^{1/2}_1 \rho_D A^{1/2}_1} {\text{Tr}\left(A^{1/2}_0 \rho_D A^{1/2}_0 + A^{1/2}_1 \rho_D A^{1/2}_1 \right)} \nonumber \\
   &=& \gamma_0 |d^\perp_{123}\rangle\langle d^\perp_{123}| + \gamma_1 |d^\perp_{023}\rangle\langle d^\perp_{023}|, 
   \end{eqnarray*}
   where $\gamma_0+\gamma_1=1$ and \begin{equation}
   \gamma_0=\frac{|\alpha|^2|\langle d^\perp_{123}|d_0\rangle|^2}{|\alpha|^2|\langle d^\perp_{123}|d_0\rangle|^2 + |\beta|^2|\langle d^\perp_{023}|d_1\rangle|^2}.
   \end{equation}
   Note that the detector states, corresponding to those checking for passage through the first and second slits have become $|d^\perp_{123}\rangle$ and $|d^\perp_{023}\rangle$ respectively, and now the task is to obtain the distinguishability, $\overline{D}$, among them, when they are given with probabilities $\gamma_0$ and $\gamma_1$. The precise quantity is obtained by using Eq. \eqref{kaise}:
   $$ \overline{D}=1-2\sqrt{\gamma_0\gamma_1}|\langle|d^\perp_{123}|d^\perp_{023}\rangle|.$$ 
   Now since the outcomes corresponding to $A_?$ were discarded, a corresponding compensation is in order, and this is achieved by multiplying the probability, $1-\text{Tr}(A_?\rho_D)$, of not being discarded, to $\overline{D}$, to obtain the actual distinguishability, $D$, between the paths followed by the quantum system, \(Q\):
   \begin{equation}
	D=(1-\text{Tr}(A_?\rho_D)) \overline{D}. \label{eq3}
   \end{equation}
	This is the quantity that we will use to measure 
	the particle nature of the quantum system, \(Q\).
   
   \subsection{Complementarity} \label{3}
   The final step 
   is to establish a complementarity between the two quantities conceptualized in the two preceding subsections, viz. quantum coherence in an incomplete basis and path distinguishability, in the interferometric set-up considered. The maximum value of the path distinguishability as also the maximum value of the quantum coherence, over arbitrary quantum states, are equal to unity. To check the complementarity, we then  find the optimal value of $C_{tr}+D$,
   using a numerical nonlinear optimization procedure.
   We determined the optimal value of $C_{tr}+D$ to be 1.400 (converged up to the third decimal point), so that we have that for arbitrary quantum states in the interferometric set-up considered,
   \begin{equation}
   C_{tr}+D \leq 1.4,
   \end{equation}
while the individual terms on the left-hand-side, when separately maximized over all quantum states, adds to 2. 
  
 
   \section{Conclusion} \label{sec4}

  In summary, we have developed a resource theory of quantum coherence with respect to an incomplete basis of the corresponding Hilbert space.
An incomplete basis arises in several physical situations, including in 
multi-slit interference experiments of which the observer has only a partial control. 
In developing the resource theory, we began by characterizing the set of incoherent states. We subsequently defined the set of incoherent operations, and identified two classes within that set. We then proposed two families of measures to quantify quantum coherence with respect to an incomplete basis, viz. the distance-based measures and the minimal-completion measures. While the first family is widely used in resource theories, including in those of entanglement and of quantum coherence with respect to complete bases, the second one is unique to the resource theory of quantum coherence for incomplete bases. 
We then considered 
a four-slit interference setting, where only two of the slits are in control of the observer. This is a situation where the wave nature of a quantum system can naturally be estimated by the quantum coherence of its state with respect to an incomplete basis, viz. the set formed by the two states corresponding to the two slits in control of the observer. 
We then found that there exists  a complementary relation between the so-defined quantum coherence of the system and the which-path information about the paths corresponding to the two slits in control of the observer. 
   

\section*{Acknowledgments}
We acknowledge the cluster facility of the Harish-Chandra Research Institute for the numerical computations performed therein. The research of CS was partly supported  by the INFOSYS scholarship.
We acknowledge support from the Department of Science and Technology,
Government of India through the QuEST grant (grant number
DST/ICPS/QUST/Theme-3/2019/120).

\end{document}